\documentclass[final,5p,times]{elsarticle}
\usepackage{amssymb}
\usepackage{graphicx}

\begin{document}

\begin{frontmatter}

%% Title, authors and addresses

%% use the tnoteref command within \title for footnotes;
%% use the tnotetext command for theassociated footnote;
%% use the fnref command within \author or \address for footnotes;
%% use the fntext command for theassociated footnote;
%% use the corref command within \author for corresponding author footnotes;
%% use the cortext command for theassociated footnote;
%% use the ead command for the email address,
%% and the form \ead[url] for the home page:
%% \title{Title\tnoteref{label1}}
%% \tnotetext[label1]{}
%% \author{Name\corref{cor1}\fnref{label2}}
%% \ead{email address}
%% \ead[url]{home page}
%% \fntext[label2]{}
%% \cortext[cor1]{}
%% \address{Address\fnref{label3}}
%% \fntext[label3]{}

\title{Sudden death and birth of entanglement beyond the Markovian
approximation}

%% use optional labels to link authors explicitly to addresses:
%% \author[label1,label2]{}
%% \address[label1]{}
%% \address[label2]{}

\author[ad1,ad2]{Z. Y. Xu}
\author[ad1]{M. Feng\corref{cor1}}
\ead{mangfeng@wipm.ac.cn} \cortext[cor1]{Corresponding author. Tel.:
011+8627-8719-9580}

\address[ad1]{State Key Laboratory of Magnetic Resonance and Atomic and Molecular
Physics, Wuhan Institute of Physics and Mathematics, Chinese Academy
of Sciences, Wuhan 430071, China}
\address[ad2]{Graduate School of the Chinese Academy of Sciences, Beijing 100049, China}

\begin{abstract}
We investigate the entanglement dynamics of two initially entangled
qubits interacting independently with two uncorrelated reservoirs
beyond the Markovian approximation. Quite different from the
Markovian reservoirs [C. E. L\'{o}pez \textit{et al}., Phys. Rev.
Lett. 101 (2008) 080503], we find that entanglement sudden birth
(ESB) of the two reservoirs occurs without certain symmetry with
respect to the entanglement sudden death (ESD) of the two qubits. A
phenomenological interpretation of entanglement revival is also
given.
\end{abstract}

\begin{keyword}
Entanglement dynamics, Non-Markovian reservoir, Entanglement sudden
death (ESD), Entanglement sudden birth (ESB) \PACS 03.67.Mn,
03.65.Ud, 03.65.Yz
%% keywords here, in the form: keyword \sep keyword

%% PACS codes here, in the form: \PACS code \sep code

%% MSC codes here, in the form: \MSC code \sep code
%% or \MSC[2008] code \sep code (2000 is the default)

\end{keyword}

\end{frontmatter}
%% \linenumbers

%% main text

\section{Introduction}

%\label{}

%% The Appendices part is started with the command \appendix;
%% appendix sections are then done as normal sections
%% \appendix

%% \section{}
%% \label{}
The entanglement dynamics of open quantum systems has attracted
considerable interests over recent years \cite{review}. By
investigating a system of two qubits interacting independently with
their corresponding vacuum reservoirs, Ref. \cite{yu-ESD} has
pointed out the sudden termination of the entanglement initially
owned by the two qubits at a finite interval, which is called
entanglement sudden death (ESD). This intriguing phenomenon has
recently
been demonstrated experimentally by linear optics systems \cite%
{exp-lo1,exp-lo2} and by atomic ensembles \cite{exp-ae}. A great deal of
theoretical investigations of ESD have so far been reported \cite%
{yu1,Lopez,Ikram,Asma,Davidovich,Ficek1,Ficek2,Bellomo1,Bellomo2,zeno,
Bellomo3,Paris,Cao,Wang}. For example, based on the same model as in
Ref. \cite{yu-ESD}, it was found that ESD regarding the qubits would
always occur under the thermal and squeezing reservoirs
\cite{Ikram,Asma}. Meanwhile,
studies of ESD have been extended to multi-particle systems \cite{Davidovich}%
, which indicated the delay of ESD time with more qubits involved.
In addition, some studies have also showed that entanglement of
qubits will revive in the case of a commonly shared reservoir
\footnote{We are aware of two very recent papers: Phys. Rev. A 79
(2009) 012301 and Phys. Rev. A 79 (2009) 042302 which addressed
similar topics in entanglement revival of two qubits shared with a
common reservoir.} \cite{Ficek1,Ficek2} or of independent
non-Markovian reservoirs \cite{Bellomo1,Bellomo2,zeno}.

A question naturally arises: where has the lost entanglement gone
when ESD occurs? An answer to the question has been given recently
by \cite{Lopez}. Considering the systems and the reservoirs as a
whole, Ref. \cite{Lopez} presented the lost entanglement to be
transferred to reservoir degrees of freedom, which is called
"entanglement sudden birth" (ESB) of the reservoirs. Intriguingly,
ESD of the systems and ESB of the reservoirs are of some symmetry
and ESB might be occurring before, simultaneously with or even after
ESD, depending on different initial states. However, the discussions
in \cite{Lopez} are restricted to the weak coupling regime under the
Markovian approximation, which only works when the reservoir
correlation
time is small compared to the relaxation time of the system \cite{open-book}%
. If the qubit is strongly coupled to its environment or the characteristic
time of the system is shorter than the environmental correlation time \cite%
{open-book,solid}, such as in recent experiments with cavity QED system \cite%
{exp-QED} or solid-state systems \cite{exp-solid,exp-spin}, the
situation would be much complicated and we have to give up the
Markovian approximation in treatment.

The present paper is focused on the study of entanglement dynamics
without the Markovian approximation for two qubits interacting,
respectively, with two independent reservoirs. For the initial
condition with the two qubits entangled, we will study ESD and ESB,
and show their different variation from the Markovian treatment.

\section{Physical model}

We consider a two-qubit system coupling to two uncorrelated vacuum
reservoirs at zero temperature. Since there is no interaction between the
two pairs of qubit-reservoir, the dynamics of the whole system can be
obtained simply from the evolution of the individual pairs \cite%
{Lopez,Bellomo1}. The Hamiltonian of the individual qubit-reservoir pair
under the rotating wave approximation is given by ($\hbar =1),$%
\begin{equation}
H=\omega _{0}\sigma _{+}\sigma _{-}+\sum_{k=1}^{N}\omega
_{k}a_{k}^{+}a_{k}+\sum_{k=1}^{N}g_{k}(\sigma _{-}a_{k}^{+}+\sigma
_{+}a_{k}),  \label{1}
\end{equation}%
where $\omega _{0}$ is the resonant transition frequency of the qubit
between levels $|0\rangle $ and $|1\rangle ,$ with $\sigma _{+}=|1\rangle
\left\langle 0\right\vert $ and $\sigma _{-}=|0\rangle \left\langle
1\right\vert $, $\omega _{k},$ $a_{k}^{+}$ $(a_{k}),$ $g_{k}$ are the
frequency, creation (annihilation) operator and the coupling constant for
the \textit{k}th mode of the reservoir. For simplicity in our treatment,
only one excitation of the reservoir will be considered. Consequently, the
initial state $|\Psi \left( 0\right) \rangle =C_{0}(0)|1\rangle _{q}|\tilde{0%
}\rangle _{r}+\sum_{k=1}^{N}C_{k}(0)|0\rangle _{q}|1_{k}\rangle _{r}$ will
evolve to

\begin{equation}
|\Psi\left( t\right) \rangle=C_{0}(t)|1\rangle_{q}|\tilde{0}%
\rangle_{r}+\sum_{k=1}^{N}C_{k}(t)|0\rangle_{q}|1_{k}\rangle_{r},  \label{2}
\end{equation}
where $|\tilde{0}\rangle_{r}=\bigotimes_{k=1}^{N}|0_{k}\rangle_{r},$ and $%
|1_{k}\rangle_{r},$ an abbreviation of $\bigotimes_{j=1,j\neq
k}^{N}|0_{j}\rangle_{r}|1_{k}\rangle_{r},$ is the reservoir state with one
excitation in the \textit{k}th mode and other states in vacuum. Assuming the
initial conditions $C_{0}(0)=1$ and $C_{k}(0)=0,$ we could obtain the
following equation by solving the Schr\"{o}dinger equation of Eq. (2)$,$%
\begin{equation}
\dot{C}_{0}(t)=-\int\limits_{0}^{t}dt^{^{\prime}}F(t-t^{^{%
\prime}})C_{0}(t^{^{\prime}}),  \label{3}
\end{equation}
where the correlation function $F(t-t^{^{\prime}})$ in the limit of $%
N\rightarrow\infty$ is of the form $F(t-t^{^{\prime}})=\int d\omega
J(\omega)e^{i(\omega_{0}-\omega)(t-t^{^{\prime}})}$ with $J(\omega)$ the
spectral density of the reservoir. In what follows, we will consider in our
calculation the Lorentzian spectral distribution $J(\omega)=\frac{\gamma_{0}%
}{2\pi}\frac{\gamma^{2}}{(\omega_{0}-\omega)^{2}+\gamma^{2}}$ \cite%
{open-book}$,$ in which $\gamma_{0}\simeq\tau_{0}^{-1}$ and $%
\gamma\simeq\tau_{r}^{-1},$ with $\tau_{0}$ and $\tau_{r}$ the qubit
relaxation time and the reservoir correlation time, respectively \cite%
{open-book}. In a non-Markovian regime, $\gamma_{0}>\gamma/2,$ i.e.,
the reservoir correlation time is longer than the relaxation time of
the qubit, and $\gamma_{0}<\gamma/2$ means a Markovian regime
\cite{open-book}$.$ As a result, in the non-Markovian regime, the
probability amplitude can be easily
solved as%
\begin{equation}
C_{0}(t)=e^{-\frac{\gamma t}{2}}\left[ \cos(\frac{\Gamma t}{2})+\frac{\gamma
}{\Gamma}\sin(\frac{\Gamma t}{2})\right] ,  \label{4}
\end{equation}
with $\Gamma=\sqrt{\gamma(2\gamma_{0}-\gamma)}.$ If we set $\tilde{C}(t)=%
\sqrt{1-C_{0}(t)^{2}}$, Eq. (2) can be rewritten as%
\begin{equation}
|\Psi\left( t\right) \rangle=C_{0}(t)|1\rangle_{q}|\tilde{0}\rangle _{r}+%
\tilde{C}(t)|0\rangle_{q}|\tilde{1}\rangle_{r},  \label{5}
\end{equation}
with $|\tilde{1}\rangle_{r}=\left[ \sum_{k=1}^{N}C_{k}(t)|1_{k}\rangle _{r}%
\right] /\tilde{C}(t)$ ($N\rightarrow\infty$).

\section{Entanglement dynamics}

We will employ concurrence \cite{Wootters} to assess entanglement. In the
case of the density matrix with the "X" type%
\begin{equation}
\rho=\left(
\begin{array}{cccc}
a & 0 & 0 & w \\
0 & b & z & 0 \\
0 & z^{\ast} & c & 0 \\
w^{\ast} & 0 & 0 & d%
\end{array}
\right) ,  \label{6}
\end{equation}
the concurrence could be easily calculated by \cite{yu-c}%
\begin{equation}
\mathbf{C}\left( \rho\right) =2\max\{0,|z|-\sqrt{ad},|w|-\sqrt{bc}\}.
\label{7}
\end{equation}

We consider that the two qubits are initially entangled and the two
reservoirs are in zero-temperature vacuum states, which is described as \cite%
{yu-ESD},%
\begin{equation}
\rho \left( 0\right) =\frac{1}{3}\left(
\begin{array}{cccc}
\alpha & 0 & 0 & 0 \\
0 & 1 & 1 & 0 \\
0 & 1 & 1 & 0 \\
0 & 0 & 0 & 1-\alpha%
\end{array}%
\right) _{q_{1}q_{2}}\otimes |\tilde{0}\rangle _{r_{1}}|\tilde{0}\rangle
_{r_{2}}.  \label{8}
\end{equation}%
Using Eq. (5), we could straightforwardly reach the reduced density matrices
for the two qubits,%
\begin{equation}
\tiny
\begin{array}{l}
\rho _{q_{1}q_{2}}(t)= \\
\frac{1}{3}\left(
\begin{array}{cccc}
\alpha C_{0}^{4}(t) & 0 & 0 & 0 \\
0 & \alpha C_{0}^{2}(t)\tilde{C}^{2}(t)+C_{0}^{2}(t) & C_{0}^{2}(t) & 0 \\
0 & C_{0}^{2}(t) & \alpha C_{0}^{2}(t)\tilde{C}^{2}(t)+C_{0}^{2}(t) & 0 \\
0 & 0 & 0 & \alpha \tilde{C}^{4}(t)+2\tilde{C}^{2}(t)+1-\alpha%
\end{array}%
\right) ,%
\end{array}
\label{9}
\end{equation}%
for the two reservoirs,%
\begin{equation}
\tiny
\begin{array}{l}
\rho _{r_{1}r_{2}}(t)= \\
\frac{1}{3}\left(
\begin{array}{cccc}
\alpha \tilde{C}^{4}(t) & 0 & 0 & 0 \\
0 & \alpha C_{0}^{2}(t)\tilde{C}^{2}(t)+\tilde{C}^{2}(t) & \tilde{C}^{2}(t)
& 0 \\
0 & \tilde{C}^{2}(t) & \alpha C_{0}^{2}(t)\tilde{C}^{2}(t)+\tilde{C}^{2}(t)
& 0 \\
0 & 0 & 0 & \alpha C_{0}^{4}(t)+2C_{0}^{2}(t)+1-\alpha%
\end{array}%
\right) ,%
\end{array}
\label{10}
\end{equation}%
for the qubit-1 and reservoir-1,%
\begin{equation}
\tiny
\begin{array}{l}
\rho _{q_{1}r_{1}}(t)= \\
\frac{1}{3}\left(
\begin{array}{cccc}
0 & 0 & 0 & 0 \\
0 & (1+\alpha )C_{0}^{2}(t) & (1+\alpha )C_{0}(t)\tilde{C}(t) & 0 \\
0 & (1+\alpha )C_{0}(t)\tilde{C}(t) & (1+\alpha )\tilde{C}^{2}(t) & 0 \\
0 & 0 & 0 & 2-\alpha
\end{array}%
\right) ,%
\end{array}
\label{11}
\end{equation}%
and for the qubit-1 and reservoir-2,%
\begin{equation}
\tiny
\begin{array}{l}
\rho _{q_{1}r_{2}}(t)= \\
\frac{1}{3}\left(
\begin{array}{cccc}
\alpha C_{0}^{2}(t)\tilde{C}^{2}(t) & 0 & 0 & 0 \\
0 & C_{0}^{2}(t)+\alpha C_{0}^{4}(t) & C_{0}(t)\tilde{C}(t) & 0 \\
0 & C_{0}(t)\tilde{C}(t) & \tilde{C}^{2}(t)+\alpha \tilde{C}^{4}(t) & 0 \\
0 & 0 & 0 & 2-\alpha +\alpha C_{0}^{2}(t)\tilde{C}^{2}(t)%
\end{array}%
\right) .%
\end{array}
\label{12}
\end{equation}%

As Eqs. (9)-(12) are all of the "X" types, the concurrence of the partitions
$q_{1}-q_{2},$ $r_{1}-r_{2}$, $q_{1}-r_{1}$ and $q_{1}-r_{2}$ can be
calculated directly using Eq. (7). In addition, Eq. (7) also implies that
the ESD time of the qubits or the ESB time of the reservoirs can be obtained
from $|z|-\sqrt{ad}=0$ and $|w|-\sqrt{bc}=0.$ However, as we could not solve
them in an explicitly analytical way, numerical calculations are given below
instead.

\section{Numerical results and discussions}

\begin{figure*}
\centering
\includegraphics[width=18.5cm]{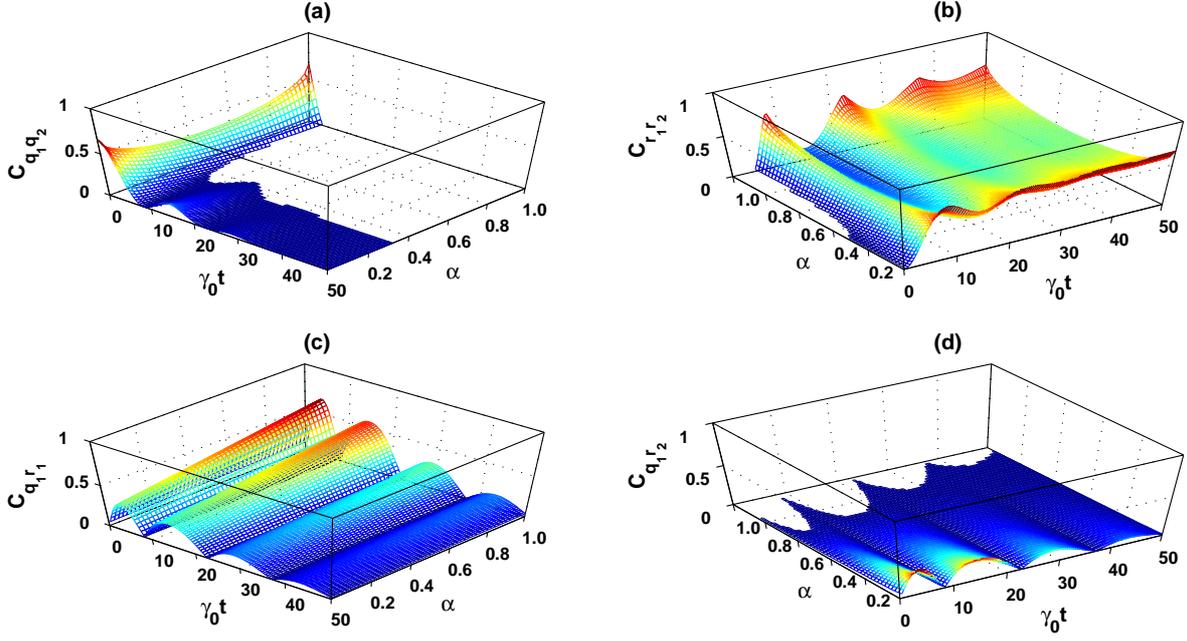}% Here is how to import EPS art
\caption{(Color online) Concurrence as a function of $\protect\alpha
$ and $\protect\gamma _{0}t$ for Eqs. (9)-(12) in the non-Markovian
regime with $\protect\gamma =0.1\protect\gamma _{0}$.}
\label{fig:wide}
\end{figure*}

Fig. 1 shows the entanglement dynamics of $q_{1}-q_{2}$, $r_{1}-r_{2}$, $%
q_{1}-r_{1},$ and $q_{1}-r_{2}$ in the non-Markovian regime with
$\gamma =0.1\gamma _{0}.$ In Fig. 1(a), there are three regimes for
entanglement variance regarding the two qubits: for $\alpha \lesssim
0.3,$ no ESD will occur; for $0.3\lesssim \alpha \lesssim 0.4$ there
will be entanglement revival after ESD takes place; as for $\alpha
\gtrsim 0.4$ entanglement will vanish forever at a finite interval
\cite{Bellomo1}. In contrast to the behavior regarding the two
qubits, there are only two $\alpha -$dependent regimes for
entanglement variance regarding the two reservoirs: ESB or no ESB,
as shown in Fig. 1 (b). But no matter in which regime, the two
reservoirs will finally be entangled in a time-dependently
oscillating manner. The entanglement transfer \footnote{To avoid
misunderstanding, we would like to mention that the term
"entanglement transfer" used in the present paper means the
appearance of a lost entanglement in some other degrees of freedom,
e.g., qubit-reservoir, reservoir-reservoir etc. due to complicated
couplings between the qubits and the environment. This is quite
different from in Refs.
\cite{trans1,trans2,trans3,trans4,trans5,trans6} etc., where
"entanglement transfer" is referred to the movement of entanglement
from the flying qubits to static qubits by some operations.} is done
step by step from the qubits to the reservoirs, which is reflected
in Figs. 1(c) and 1(d) with the entanglement variance regarding the
partitions $q_{1}-r_{1}$, $q_{1}-r_{2}$ etc. . The partitions
$q_{1}-r_{1}$, $q_{1}-r_{2}$ behave as something like relays, which
get and release entanglement in an oscillating way. Meanwhile the
$r_{1}-r_{2}$ entanglement is enhanced also in the oscillating way.

\begin{figure*}
\centering
\includegraphics[width=14cm]{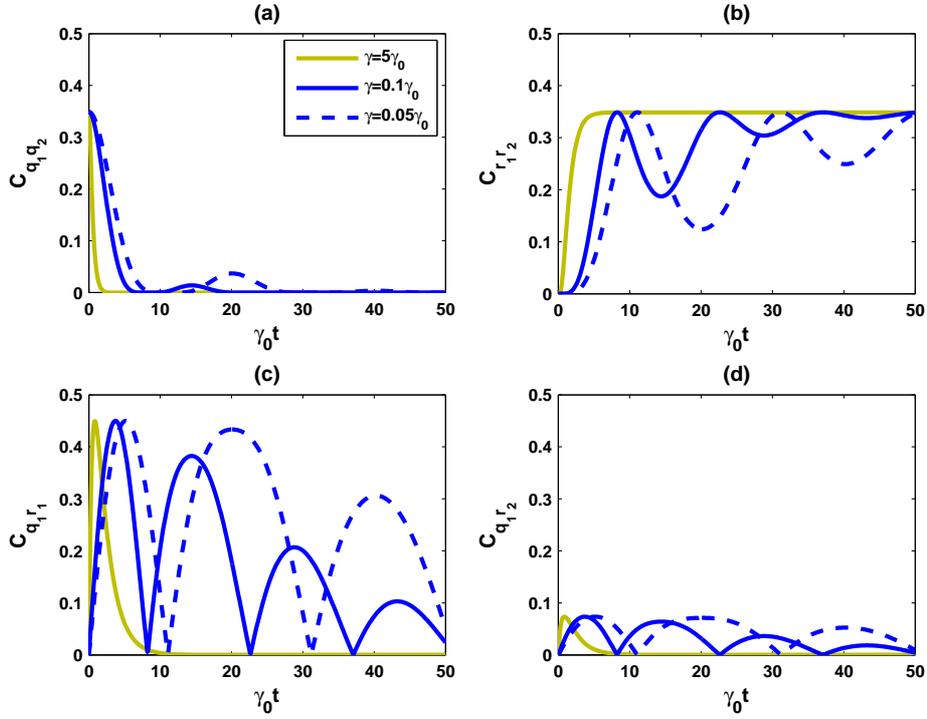}% Here is how to import EPS art
\caption{(Color online) Concurrence as a function of $\protect\gamma
_{0}t$ with $\protect\alpha =0.35$ for Eqs. (9)-(12) in different
coupling
intensities where the green (grey) solid lines mean the Markovian case with $%
\protect\gamma =5\protect\gamma _{0};$ the blue curves correspond to
the non-Markovian cases with $\protect\gamma =0.1\protect\gamma
_{0}$ (solid lines) and $\protect\gamma =0.05\protect\gamma _{0}$
(dashed lines) respectively.}
\end{figure*}

To see more clearly and also to make comparison with the Markovian
treatment, we have plotted Fig. 2 for the entanglement dynamics with
$\alpha
=0.35$ $\left( t_{ESD}>t_{ESB}\right) .$ In the non-Markovian regime, e.g., $%
\gamma =0.05\gamma _{0},$ when ESD occurs at $\gamma _{0}t\simeq 8$, the
entanglement of the qubits has been transferred to other partitions, e.g., $%
r_{1}-r_{2}$, $q_{1}-r_{1},$ $q_{1}-r_{2}$ etc. . In more details,
we may first visualize that within an interval $\gamma _{0}t\in
\lbrack 0,5.5),$ in which the entanglement transfer from qubit-qubit
to qubit-reservoir is greater than that from qubit-reservoir to
reservoir-reservoir degrees of freedom, and the entanglement of
qubit-reservoir and reservoir-reservoir are both increasing. Then
from the critical point $\gamma _{0}t\simeq 5.5,$ the
qubit-reservoir entanglement begins to be decreasing until $\gamma
_{0}t\simeq 11,$ at which the entanglement of qubit-reservoir has
been totally transferred to the reservoir degrees of freedom.
However, in contrast to the case of Markovian reservoirs with the
entanglement fully transferred to the reservoirs at one time, due to
the memory effect of the non-Markovian reservoirs, entanglement will
partially be recoiled to the qubit-reservoir degrees of freedom, or
even back to the qubit-qubit degrees
of freedom. As a result, revival of entanglement takes place \cite%
{Bellomo1,Bellomo2}. This back and forth process will repeat until
the entanglement of the qubits have been completely moved to their
corresponding reservoirs. Finally, the reservoirs are entangled.
Moreover, Fig. 2 also presents that, the stronger the qubits
coupling to their reservoirs, the more evident the non-Markovian
memory effects.

\section{Conclusions}

In summary,\textbf{\ }we have investigated two-qubit entanglement affected
by the decoherence from non-Markovian reservoirs. In contrast to the smooth
decay of the qubit entanglement in the case of Markovian reservoirs, the
entanglement of the qubits will revive after ESD occurs in the non-Markovian
case. Meanwhile, by means of the qubit-reservoir partitions as the relays,
the reservoirs will be asymptotically entangled in an oscillating manner,
which is a good interpretation of entanglement revival of qubits in the
non-Markovian reservoirs. We argue that our present study would be useful
for quantum information science. For example, the solid-state qubits in
systems with strong correlation always experience strong decoherence \cite%
{solid,exp-solid,exp-spin}. So quantum information processing on any qubit
in such systems should be affected by detrimental influence from the
non-Markovian reservoir. Moreover, fast logic gating is necessary in quantum
information processing. As the characteristic time of the system during the
fast gating is very short, we have to consider the non-Markovian effect from
environment in assessing the operational fidelity. We have noticed recent
studies on non-Markovian reservoir from the microscopic way \cite%
{zhang1,zhang2}. Our present work from the phenomenological viewpoint,
however, could explain some experimental observations of entanglement loss
regarding qubits more straightforwardly. On the other hand, entanglement is
the unique resource in quantum information science. As no system could be
completely isolated from external noise, it is of interest to understand
what happens in the entanglement transfer from the system to the
environment, although whether the reservoirs are entangled or not is
actually of no practical application with current technology.

\section*{Acknowledgements}

The work is supported by NNSF of China under Grants No. 10774163 and No.
10774042.

\end{document}